# Quantum Matter-Photonics Framework: Analyses of Chemical Conversion Processes

O. Tapia

A quantum photonics framework is adapted to help scrutinize chemical reaction mechanisms and used to explore a process mapped from $C\text{-}(H)_2C(H)OH \Leftrightarrow C(H)_3\text{-}CHO$ topologic model. The chemical concept of bond knitting/breaking is reformulated via partitioned base sets leading to an abstract and general quantum presentation. Pivotal roles are assigned to entanglement, coherence, decoherence, and Feshbach resonance quantum states that permit apprehend gating states in conversion processes. A view from "above" in the state energy eigenvalue ladder, belonging to electro-nuclear spectra complement the standard view from ground state. A full quantum physical view supporting chemical change obtains that adapts to high-energy probes such as laser pulses yet naturally letting in external low frequency as "chemical" drivers (catalysts).

**Introduction**

A photonic framework including an abstract multi-partite base set opens possibilities to simultaneously handle matter-sustained and electromagnetic-sustained quantum base states. [1,2] A quantum scheme obtains by projecting abstract Hilbert space vectors on a configuration space covering classic and quantum degrees of freedom; chemical mechanisms become formalized and analyzed within a full quantum physical scheme; early developments leading to present work are found in refs. [3-13]

Progresses in quantum technology [14-20] are producing important results making available varied laser sources as well as probing and modulating devices with which to drive chemical processes starting from high energy levels thereby opening new vistas; pump-probe experiments illustrate the issue. These advances sooner or later will influence the way we quantum chemists look at mechanisms sensed to describe reacting chemical systems; e.g. use of molecular states to encode quantum information, [21,22] or starting processes well above standard ionization and/or dissociation limit states; attosecond pulses and subsequent modulation with low frequency electromagnetic (EM) radiation acting as effectors display a different character of quantum mechanisms; as a matter of fact, they require as a particular case external microwaves.

Thus far, standard mechanistic descriptions of electronuclear (EN) systems work with fixed nuclei and electronic spin states while present-day requirements from experiments would rather involve varieties of angular momenta multiplicities and no nuclei localizations; see e.g.

[16] Mixing of base states with different total angular momenta entails inclusion e.g. of spin-orbit and other coupling operators to follow responses rooted at quantum base states ranging from reactants to products. Moreover, explicit consideration of quantized EM fields becomes necessary as the bosonic character, spin S=1, enters in the rules of angular momentum conservation. This situation compels a level of theory beyond that one underlying standard electronic quantum chemical technology, see ref. [23] A call is made for grounding of present computing schemes beyond classical mechanically flavored representations; accordingly, implementing matter-photonics schemes [1] aiming at an enhancement of the scope of our quantum chemical framework becomes a must. Thereafter, wave functions, projections of abstract elements (abstract Hilbert space) over a configuration space, can be charted into the domain of physical wave functions; [5,24] but now, materiality sustains such q-states thereby overcoming standard representation scheme grounding molecular models. It is the field of possibilities that opens as ground for the present approach.

Target sought: relate probes measurements to physical wave functions and vice-versa. [5,24] For a quantum chemist this might be a novel situation but this imposes on us a different view. [1-3,25-29] From our perspective, a q-state no longer represents matter elementary constituents as if they were objects (e.g. molecules); neither as particles occupying energy levels. [26-29]

Grasping phenomena signaled above [14-21] requires inclusion of photon field angular momentum and a fair amount of abstract quantum physical concepts [3,25] not yet available in standard computational-chemistry; the standard technology, based on BO adiabatic models, [30] includes concepts of classical potential energy surfaces (PES) that may conceal difficulties when handling mechanistic studies of so called bond break/forming as shown by Crespo *et al*. [25] The present paper, combined to refs [26-28] starts up assuming diabatic perspectives [4] for the *matter sustained* q-states and move forward the mathematical formalism designed to help grip electronuclear non-separability from quantum numbers perspective. [3-5,23] Thereafter, different levels of entanglement between photon and matter base states appear in the theory. In principle, partial separability by quantum numbers provides chemically flavored basis set. [3,4] What is still lacking is a bridge with actual advanced multi-configuration calculations that would help practitioners fit in numerical results into a richer quantum physical perspective. Efforts to bridge this gap are initiated here.

The topology of reaction system $CH_2=CHOH \Leftrightarrow CH_3-HC=O$ provides simple elements to examine keto-enol conversion that would help mapping chemical concepts of bond-reshuffling processes and concomitant organic function exchanges in the context now of abstract quantum physical schemes. [31] The theory so obtained within the photonic framework would permit assessing symmetries in EN space; also, all possible EN excited states sustained by the elementary material elements, common to keto and enol forms, [3,4] would enter via quantum numbers embedded in abstract multi partite bases framework; here, chemical formulae take on topologic scope. [31]

Matters are organized as follows. Section 2 situates the problem; it starts identifying relevant stationary partitioned (partite) EN states: Sect.2.1 introduces multi-partite base sets; Sect.2.2 presents inertial frames required to define configuration spaces. A first connection between laboratory and abstract spaces is established. Sect.2.3 indicates the joint between semi-classic matter-radiation probing formalism. Section 2.4 focuses on formal photonic framework definition; the basis vector would include all possibilities the system may show under quantum probing conditions, the line vector over basis states defines possibilities accessible to the system; the column vector including all complex amplitudes retains all those zero entries that complete the q-state. It is up to the experimenter to define apparatuses located in real space that are able to interact with the physical q-state to get responses that, in turn, are to be used in characterizing it.

Section 3 present quantum base states for Keto/Enol chemical processes; information over such states could be gathered from conveniently adapted quantum-chemical models. Section 3.1 includes a discussion of Keto/Enol mechanism viewed from this new photonic perspective. And 3.2 presents a rounding off by including external sources that will drive the system via amplitudes changes. The role of I-frames is emphasized.

Section 4 closes with a set of general remarks. A physical picture emerges that reminds on the one hand, the role of Feshbach quantum states for material systems entangled with the radiation field. These resonances would act as gates to changing amplitudes at entangled bound base states; or as exit towards entangled asymptotically separated molecular states, thereby emphasizing the quantum physical nature of chemical bond reshuffling processes.

## 2. Theory: Elements

To start positing the problem, consider first an abstract multipartite scheme adapted to examine chemical and physical processes. [1,2,26-28] Begin with sets of $k'_1 \ldots k'_m$ as labels (quantum numbers) for basis kets partitioned as:

| | | |
|---|---|---|
| 1) | $|k'_1 \ldots k'_m\rangle$ | (2;1) |
| 2a) | $|k'_1 \ldots k'_{m-1}\rangle \otimes |k'_m\rangle$ | (2;2a) |
| 2b) | $|k'_1\rangle \otimes |k'_2 \ldots k'_m\rangle$ | (2;2b) |
| | …. | |
| 3a) | $|k'_1 \ldots k'_{m-2}\rangle \otimes |k'_{m-1}\rangle \otimes |k'_m\rangle$ | (2;3a) |
| 3b) | $|k'_1\rangle \otimes |k'_2 \ldots k'_{m-1}\rangle \otimes |k'_m\rangle$ | (2;3b) |
| | …. | |
| m) | $|k'_1\rangle \otimes |k'_2\rangle \otimes \ldots \otimes |k'_{m-1}\rangle \otimes |k'_m\rangle$ | (2;m) |

These abstract states are projected over configuration spaces as indicated now.

### 2.1. Projected multi-partite basis sets

Include a configuration space. Its dimension embodies the number 3m of classic degrees of freedom: $(\mathbf{x}_1, \ldots, \mathbf{x}_m)$; the information mediates a relation between abstract states and projected ones, namely a wave function. A form of generic basis sets given with the help of quantum number sets reads as:

1)        $\langle \mathbf{x}_1,\ldots,\mathbf{x}_m | \phi_{k'1\ldots k'm} \rangle$        (2.1;1)

2a)      $\langle \mathbf{x}_1,\ldots,\mathbf{x}_{m-1} | \phi_{k'1\ldots k'm-1} \rangle \otimes \langle \mathbf{x}_m | \phi_{k'm} \rangle$        (2.1;2a)

2b)      $\langle \mathbf{x}_1 | \phi_{k'1} \rangle \otimes \langle \mathbf{x}_2,\ldots,\mathbf{x}_{m-1},\mathbf{x}_m | \phi_{k'2\ldots k'm-1\ k'm} \rangle$        (2.1;2b)

2c)      $\langle \mathbf{x}_1,\ldots \mathbf{x}_v | \phi_{k'1\ldots \phi_{k'v}} \rangle \otimes_v \langle \mathbf{x}_{v+1},\ldots,\mathbf{x}_m | \phi_{k'v+1\ldots k'm} \rangle$        (2.1;2c)

           ….

3a)      $\langle \mathbf{x}_1,\ldots,\mathbf{x}_{m-2} | \phi_{k'1\ldots k'm-2} \rangle \otimes_{m-2} \langle \mathbf{x}_{m-1} | \phi_{k'm-1} \rangle \otimes_{m-1} \langle \mathbf{x}_m | \phi_{k'm} \rangle$        (2.1;3a)

3b)      $\langle \mathbf{x}_1 | \phi_{k'1} \rangle \otimes_1 \langle \mathbf{x}_2,\ldots,\mathbf{x}_{m-1} | \phi_{k'2\ldots k'm-1} \rangle \otimes_{m-1} \langle \mathbf{x}_m | \phi_{k'm} \rangle$        (2.1;3b)

           ….

m)      $\langle \mathbf{x}_1 | \phi_{k'1} \rangle \otimes_1 \langle \mathbf{x}_2 | \phi_{k'2} \rangle \otimes_2 \ldots \otimes_{m-2} \langle \mathbf{x}_{m-1} | \phi_{k'm-1} \rangle \otimes_{m-1} \langle \mathbf{x}_m | \phi_{k'm} \rangle$    (2.1;m)

The symbol $\otimes$ introduces one nodal plane view from an I-frame perspective. Intended situations can briefly be given as follows:

Sector 1) stands for all possible one-partite base states; e.g. the lowest energy arrangement mapped to an EN (super)-molecular ground state. Here, it is only an abstract base state label with energy values not yet indicated. For example, keto/enol would correspond to two closed shell lowest energy root states; the chemical functionality permits also naming base states (see below). [31] But for a moment let us keep the abstract presentation.

Sector 2) corresponds to bi-partite states; selecting three illustrative cases a, b, c that, extrapolated to chemistry might be akin to dissociated fragment states.

Sector 3) show examples of tri-partite base set. These stand e.g. for two no-bonding fragment and one set of core states.

Sector m) corresponds to m-partite base sets; in a physical chemical framework would be akin to m-independent "components".

These sectors render large flexibility where the finite number of classical degrees of freedom associates to infinite number of basis functions; these latter stand for all accessible possibilities open to a constant materiality. Distinction between hadronic and leptonic degrees of freedom would facilitate presentation of ionic states later on.

The quantum physical study of a materiality is given in terms of 3m degrees of freedom (dof). Inclusion of quantum dof basis sets implies all elements signaled by eqs.(2.1;1) up to (2.1;m). Note: Incorporated into abstract space referred to a unique I-frame the global multipartite basis set may sustain coherent quantum states, formally given as:

$$|\Xi\rangle = \sum_{\text{over-all-sectors}} C_{|(2.1;1)\rangle}(\Xi)\ |(2.1;1)\rangle + \ldots + C_{|(2.1;m)\rangle}(\Xi)\ |(2.1;m)\rangle \qquad (2.1.2)$$

A quantum state determines all response *possibilities* towards external probes via definite sets of amplitudes. Different decoherence levels may associate partite base states to different I-frames and consequently can be simulated with semi-classical models. In particular, consider the following linear superposition forms:

$$|\kappa\rangle = \Sigma_{k'1\ldots k'm}\, C_{k'1k'2\ldots k'm-1k'm}(\kappa)\, \langle x_1,\ldots,x_m|\phi_{k'1\ldots k'm}\rangle + \ldots +$$
$$C_{k'1\otimes k'2\otimes\ldots\otimes k'm-1\cdots\otimes k'm}(\kappa)\, \langle x_1|\phi_{k'1}\rangle\otimes\langle x_2|\phi_{k'2}\rangle\otimes\ldots\otimes\langle x_{m-1}|\phi_{k'm-1}\rangle\otimes\langle x_m|\phi_{k'm}\rangle \rightarrow$$

$$(\langle x_1,\ldots,x_m|\phi_{k'1\ldots k'm}\rangle \ldots \langle x_1|\phi_{k'1}\rangle\otimes\langle x_2|\phi_{k'2}\rangle\otimes\ldots\otimes\langle x_{m-1}|\phi_{k'm-1}\rangle\otimes\langle x_m|\phi_{k'm}\rangle\ldots)\, \bullet$$
$$(C_{k'1k'2\ldots k'm-1k'm}\ldots\, C_{k'1\otimes k'2\otimes\ldots k'm-1\cdots\otimes k'm}\ldots)^T \qquad (2.1.3)$$

A state $|\kappa\rangle$ belongs to a coherent set if all partite elements amplitudes are referred to one and the same I-frame. In the case where all partite elements are described by independent I-frames the form (2.1.3) with all Cs=0 except $C_{k'1\otimes k'2\ldots\otimes k'm-1\otimes k'm} \neq 0$ reckons a fully de-coherent state (it is actually not an element of Hilbert space). Such symbol may relate for example to a gas of I-frames (particles) with internal quantum structure. Now, as soon as correlations develop for the case of finite subsets of integer spin partite the superposition may take on a coherent edge; the system may go through a phase change, signal by amplitude variations, where for example a Bose-Einstein condensate (BEC) would correspond to a global coherent form or to partially coherent ones. Thus the q-state $(\ldots 0_{k'1k'2\ldots k'm-1k'm}\ldots 1_{k'1\otimes k'2\otimes\ldots k'm-1\cdots\otimes k'm}\ldots)^T$ would stand for the least correlated state, while $(\ldots 1_{k'1k'2\ldots k'm-1k'm}\ldots 0_{k'1\otimes k'2\otimes\ldots k'm-1\cdots\otimes k'm}\ldots)^T$ would correspond to full BEC-like state, in chemical language a sort of "super-molecule". The change in amplitudes pattern signals the change in spectral responses.

Spectral responses to external probes permit distinguishing different types of q-states and partite base states thereof. Quantum degrees are given as q-numbers and mathematical forms, e.g. spinorial forms.

**2.2. Inertial frames**

The role of inertial frames, and more generally space-time frames is manifest yet seldom emphasized. An I-frame is characteristic of special relativity theory (SRT); in space-time a privileged time-direction is chosen. [32] It is at this stage that important concepts relevant to relationships between laboratory systems and quantum dimensions can be introduced.

Let M be a space-time continuum. The mapping above $\langle x_1,\ldots,x_m|\phi_{k'1\ldots k'm}\rangle = \phi_{k'1\ldots k'm}(x_1,\ldots,x_m)$ and the functions $\phi_{k'1\ldots k'm}(x)$ relates M to some vector space V over complex numbers. In a *neighborhood* of $x$ the wave function $\phi_{k'1\ldots k'm}(x)$ contains information on the abstract quantum base state $|\phi_{k'1\ldots k'm}\rangle$. This transformation to the space of complex functions over real support opens quantum physics to mathematical physics. Consider now the space of all possible reference (r) frames $P_x$. And let any two r-frames be uniquely related by an element of a transformation group G. Thus, I-frames permit introduction e.g. of translation, rotation and other symmetry operations; in particular Lorentz groups. A smooth concatenation P of various $P_x$ as $x$ ranges over M is called a principal fiber bundle with group G. A *continuous choice* of reference frame is called a *gauge*. [33] We see the import of introducing I-frames in the theoretic framework where changes of quantum states are the issue.

Note that the abstract state base, ket $|\phi_{k'1...k'm}>$, must be completely invariant to frame choice and symmetry operations. Symmetry properties can be endowed on projected elements only.

For these base sets coherent linear superpositions would stand for q-states all referred to a fix I-frame. The same number of elementary material constituents may sustain all these varied types of base functions. Now, the concepts of object and/or entity (molecule) are set aside; the quantum state concept takes over the one characteristic of chemistry: namely, chemical compounds; this latter concept is not appropriate, e.g. under the effects of attosecond pulses. See other experimental aspects in ref.[14]

Including an I-frame system in a volume V permits introduction of box quantum numbers; elementary material elements trapped in a box. The materiality sustains these base states too; box quantum numbers help construct base sets for I-frames external (global) quantum states. I-frame state labels correspond with box q-numbers. Incidentally, it is apparent that two I-frames are required to fix internal and external q-numbers; or defining symmetry operations. An EN base state $|jm(j)>$ signals first an electronic quantum number (j), and subsidiary quantum numbers m(j) related to nuclear dof; they express in coherent lattice dynamics. [34]

### 2.4. Matter-photonic framework

Fock space relates to EM fields via base vectors: $\{|n_\omega>\}$; the set $n_\omega = 0,1,...$ label the number of EM energy quanta available in a given circumstance. Thus for each color (frequency ω) the base vectors inform on *number of available* energy quanta; these symbols do not represent photons as if they were objects. [29]

The photonic bases for each color and matter-sustained base one forms a fourfold subset. [1,26-29] For 1-photon processes typical base state vector elements read:

$$(...|jm(j)>\otimes|1_\omega>...|jm(j);1_\omega>...|j'm(j');0_\omega>...|j'm(j')>\otimes|0_\omega>...) \qquad (2.4.1)$$

From these four entries, only amplitudes at $|jm(j)>\otimes|1_\omega>$ can signal possible ingoing or outgoing element via amplitudes appearing for particular base q-states. The base state $|jm(j);1_\omega>$ on the contrary indicates entanglement between the quantized EM field and the material-sustained base state; notice that it is not "absorption", it is just a particular base state. Both sectors are no longer separable though identifiable. The value of the energy level increases by the equivalent of a photon yet no free element (photon) is available.

The base state $|j'm(j');0_\omega>$ in eq. (2.4.1) indicates entanglement of material base state and vacuum state related to one color state; the information gathered by $0_\omega$ identifies a root state located below $E_{j'm(j')=0}$. Thus, the last two terms in (2.4.1) correspond to supplemental information concerning a root state not necessarily the same associated to $|jm(j)>$; this flexibility permits handling so called Λ-systems where, for example two energy levels relate via a common one located above; Cf. [1,2]. Thus, for example the label "$0_\omega$" in the base states elements |j'm(j')>⊗|0_ω> and |j'm(j');0_ω> sign an excitation path with a root state that differ from the one in eq. (2.4.1).

These are basis sets characteristic of the photonic framework. Note that |jm(j);1_ω> and |j'm(j')>⊗|0_ω> the energy levels may or may not be degenerate; the degenerate case plays an important role as seen below. [26]

Henceforth, an arbitrary quantum state over basis set (2.4.1) is given as amplitude vector:

$$(\ldots C_{jm(j)\otimes 1_\omega}\ldots C_{jm(j);1_\omega}\ldots C_{j'm(j');0_\omega}\ldots C_{j'm(j')\otimes 0_\omega}\ldots)^T \qquad (2.4.2)$$

These symbols reckon possibilities. As noted above, the amplitude $C_{jm(j)\otimes 1_\omega}$ focus attention to photon state either incoming or outgoing with respect to the given I-frame to which elementary materiality is assigned; these would mean possibilities accessible to the system that eventually become modulated by external effectors under quantum probing conditions.

Care must be exercised with standard linear superpositions over finite base sets. For the rigorous definition corresponds to the scalar product form:

$$(\ldots |jm(j)\rangle\otimes|1_\omega\rangle \ldots |jm(j); 1_\omega\rangle \ldots |j'm(j');0_\omega\rangle \ldots |j'm(j')\rangle\otimes|0_\omega\rangle \ldots) \bullet$$
$$(\ldots C_{jm(j)\otimes 1_\omega}\ldots C_{jm(j);1_\omega}\ldots C_{j'm(j');0_\omega}\ldots C_{j'm(j')\otimes 0_\omega}\ldots)^T \to |\text{q-state}\rangle \qquad (2.4.3)$$

A |q-state> should contain information about all zero and non-zero amplitudes and respective ordering. Such information lacks in eq.(2.1.2), thus care must be exercised when using it.

Actually, the second line of (2.4.3) defines the q-state because the base set is invariant to the extent it takes care of all possibilities associated to responses from elementary material constituents. Making effective the scalar product operation, the sum that in principle would contain infinite number of zero elements, will fail to do it explicitly (order lost); conventionally, terms multiplied by zero do not appear. On the contrary, form (2.4.3) does include possibilities accessible to the materiality that sustains q-states.

The photonic scheme can sense q-states that *emerge* if for example an external time-dependent coupling is switched on; thus taking or giving energy. It can also identify decoherence windows produced by photon-state emission or absorption from the q-EM field. Note, amplitude $C_{j'm(j')\otimes 0_\omega}$ not necessarily informs on correlated EM vacuum and excited state sustained by the materiality; it may signify planning of experimentalists not yet operational. This is a distinct advantage with respect to semi-classic models. Thus, in between an ingoing step and later outgoing photon states there is no semi-classic effect; the EM-field cannot act on the material system via electric fields, as no photon is available now, and consequently cannot "feel" the semi-classic effect of EM fields.[2,3] Entering the entangled sector is the way to handle photon-matter interactions quantum physically.

At laboratory, q-state: $(\ldots 0_{jm(j)\otimes 1_\omega}\; 0_{jm(j);1_\omega} \ldots 0_{j'm(j');0_\omega}\ldots \exp(i\mathbf{k}.\mathbf{x})1_{j'm(j')\otimes 0_\omega} \ldots)^T$ signals decoherence between photon and materiality fields if there were a loss of information about the source location. Otherwise the former field depleted in one quantum of EM-energy, the latter sustain excited state |j'm(j')> all referred to the same I-frame.

The exponential term applies to the amplitude affected to Fock base state $|0_\omega>$. Again, if at a given time the four amplitudes (made plain in (2.3.3)) are non-zero one speaks of a coherent superposition.

The transpose vectors: $(\ldots \exp(i\mathbf{k}.\mathbf{x})1_{jm(j)\otimes 1_\omega}\; 0_{jm(j);1_\omega} \ldots 0_{j'm(j');0_\omega}\; 0_{j'm(j')\otimes 0_\omega} \ldots)^T$ signal possible photon state re-emission; we use plural form because there are sets of infinite possible directions **k** and the I-frame may ensure (linear) momentum conservation.

An incoming photon state is signaled by the amplitude $\exp(-i\mathbf{k}.\mathbf{x})1_{jm(j)\otimes 1_\omega}$. The direction possibilities are of course unlimited. The experimental setup will reduce possible quantum states; in this sense QM will appear as contextual.

The photonic approach includes on the one hand, information on available q-EM energy, where root and target labels complete the basis set. On the other hand, elementary matter constituents are fixed; such materiality sustains quantum states that mediate interactions with external sources.
Note that materiality does not occupy quantum states of any sort. Classical representational mode is broken in this quantum physics approach.

### 3. Keto-Enol: Mixing information

The question is: How can enol/keto states be distinguished using characteristics associated to an I-frame?
Here, use the chemical graphs in a fully topologic manner; the tree characterizing the graph is a constant (invariant) although any internal relative distances can change at will: the tree remains. T-graphs serve as labels for basis state kets (vector names). I-frames let us introduce quantum numbers.

Consider three orthogonal planes with one point in common serving as origin for the I-frame. The (x,y)-plane ($\sigma_{xy}$) intersecting (y,z)-plane ($\sigma_{yz}$) defines a y-axis; x-axis perpendicular to $\sigma_{yz}$ and z-axis orthogonal to $\sigma_{xy}$. Assume the (z,x)-plane ($\sigma_{zx}$) cuts y-axis at the origin. "Decorating" the y-axis with two labels symmetrically disposed with respect to $\sigma_{zx}$ permits signaling carbon-carbon axis invariant to substitution of global Keto or Enol labels.

Locate now topologies for $CH_2=CHOH$ and $CH_3-C(H)=O$. Each one presents a nodal plane. Take $\sigma_{xy}$ including enol π-bond C=C while $\sigma_{yz}$ contains HC –C(H)=O; there is a nodal plane for the local C=O π-bond; bisecting the methylene group (-$CH_2$) this plane defining an anti-bonding state for an H---H motif. Disjoint four local planes associated to decorating elements –$CH_3$ and –C(H)=O and let them rotate independently for instance in a disrotatory manner; one by π/4 and the other by -π/4. Assume that at each local plane there is a nodal plane (in general) so that the associated base state would show a global Δ-symmetry. Also, local Δ-symmetry generates by including disrotatory displacement in opposite directions: each decorating center may support an L=2 angular momentum. See ref. [23] for further details.

Note that excited electronic states related to methyl motif (-$CH_3$) will show symmetric and anti-symmetric fluctuations.

Relating an abstract photonic scheme to standard quantum chemical one requires more precisions. The generic task was described in ref.[1,2,26] with some detail. Scheme I (below) describes a generic basis sets. The connection provided by chemical symbols not only facilitates a descriptive language, they can be linked to topologic graphs. Briefly: enol form presents a generic nodal-plane corresponding to the so-called π-bond engaging the C-C axis. This axis "locates" anywhere, but once chosen it partially determines the I-frame, which includes the intersection of two orthogonal planes. The alcohol function lies on this π-nodal plane.

**Enol**

$CH_2=CHOH$ (S=0) ($\xi^{(E1)}$) → $|\pi^2_{CC}, S=0\rangle$ → $|1,m(1)\rangle$     (3.3-1aE)

$C(\cdot)H_2-C(\cdot)HOH$ (S=1) ($\xi^{(E2)}$) → $|\pi\pi^*_{CC}, S=1\rangle$ → $|2,m(2)\rangle$     (3.3-1bE)

$C(\cdot)H_2-C(\cdot)HOH$ (S=0) ($\xi^{(E3)}$) → $|\pi\pi^*_{CC}, S=0\rangle$ → $|3,m(3)\rangle$     (3.3-1b'E)

$C(\cdot)H_2-C(\cdot)HO(\cdot)H(\cdot)$ (S=2)($\xi^{(E4)}$) → $|\pi\pi^*_{CC}\,\sigma\sigma^*_{OH}, S=2\rangle$ → $|4,m(4)\rangle$     (3.3.1-cE)

**Keto**

$CH_3-C(H)=O$ (S=0) ($\xi^{(K1)}$) → $|\pi^2_{CO}, S=0\rangle$ → $|5,m(5)\rangle$     (3.3-1aK)

$CH_3-C(\cdot)(H)-O(\cdot)$ (S=1) ($\xi^{(K2)}$) → $|\pi\pi^*_{CO}, S=1\rangle$ → $|6,m(6)\rangle$     (3.3-1bK)

$CH_3-C(\cdot)(H)-O(\cdot)$ (S=0) ($\xi^{(K3)}$) → $|\pi\pi^*_{CO}, S=0\rangle$ → $|7,m(7)\rangle$     (3.3-1b'K)

$(\cdot)HC(\cdot)H_2-C(\cdot)(H)-O(\cdot)$ (S=2) ($\xi^{(K4)}$) → $|\sigma\sigma^*_{CH}\,\pi\pi^*_{CO}, S=2\rangle$ → $|8,m(8)\rangle$     (3.3-1cK)

**Radical base states:** bipartite states

$C(\cdot)H_2-C(\cdot)(H)O(\cdot)(S=3/2) + H(\cdot)\,(S=1/2)(\xi^{(E5)})$
    → $|\pi\pi^*_{CC}, \sigma_O, S=3/2\rangle \otimes |H;S=1/2\rangle$ → $|9,m(9)\rangle$     (3.3-2cE)

$(\cdot)H(S=1/2) + C(\cdot)H_2-C(\cdot)(H)-O(\cdot)\,(S=3/2)\,(\xi^{(K5)})$
    → $|H;S=1/2\rangle \otimes |\pi\pi^*_{CC}, \sigma_O, S=3/2\rangle$ → $|10,m(10)\rangle$     (3.3-2cK)

**Scheme I**

The keto form presents a nodal plane involving –H-C=O fragment, besides it contains one H-atom bisecting the remaining $CH_2$ group. Both nodal planes (keto and enol) being orthogonal to each other the corresponding base functions labeled independently with the help of normal vectors to such planes; thus, the I-frame permits partitioning the space without problems. And it is not difficult to identify a generic transition state region with Δ-state symmetry.

A third plane separating the C-C feature completes geometry elements associated to the I-frame. For bipartite states such as (·)CH3⊕H(·)C=O and (·)CH2⊕(·)CHOH this plane can be seen as a nodal plane. More important, the new plane cuts out sections from the preceding planes in the sense that one can rotate them either in phase or anti phase; in chemical language: rotatory and disrotatory; Cf. ref.[26] for details.

A "reaction coordinate" corresponds to a dis-rotatory displacement of local planes associated for example to bipartite elements. If one starts up from the enol ground state form note that a $\pi/4$-angle between the normal vectors to fragment planes reads $\pi/2$, this value is used to identify broken C-C $\pi$-bond: the characteristic is then an increment in numbers of nodal planes from one ($\pi$-system) to two (d- or $\Delta$-system). Moreover, the effective change of enol into keto corresponds to a change of the tree structure; chemically achieved as a hydrogen atom swap. Completing the semi-classic picture, at $\pi/4$ the alcohol group appears to be "on its way" toward the plane keto group would occupy while the C-C $\pi$-bond breaks. No symmetry operation is involved. Only, H-swapping must be executed quantum mechanically, not in classical terms "moving" H via minimum energy profiles; let discuss this issue below.

The base states |9> and |10> show their chemical origin and therefore are not compatible. Thus, base state |9> targets possible dissociation of hydroxyl function (-O-H); base state |10> targets those for C-H. Both possible dissociation channels share partite states from C(·)H$_2$-C(·)(H)-O(·) spin state (S=3/2).

According to preceding construction, $\pi$-planes (nodal) ground states frameworks for (3.3-1aE) and (3.3-1aK) should be located on perpendicular planes. So that a semi-classic computation pulling the hydrogen atom from hydroxyl to the –CH$_2$ group on the plane violates quantum mechanical selection rules.

In principle, a chemical change can be guessed if the basis includes chemically related basis states (without changing the elementary material content), say: (…|Keto>…|Enol>..) and a generic q-state would look as: (…$C_{Keto}$…$C_{Enol}$..)$^T$; i.e. it is a coherent q-state. Thus, preparing the state (…$0_{Keto}$…$1_{Enol}$..)$^T$ a quantum transformation to state (…$1_{Keto}$…$0_{Enol}$..)$^T$ would do the chemistry at the abstract quantum physical level. When amplitudes co-exist one could get simultaneous response from both base states weighted by respective amplitudes as in eq.(2.2.2). Copenhagen interpretation [29] has no place here, as there are no objects (molecules) entering the analyses; all quantum states are *sustained* by the (same) materiality; this fresh concept permits relating abstract to laboratory physical states. Consequently, materiality and properties are separated in the present view.

What is missing then? Answer: an adequate quantum dynamical grasp of mechanisms.

### 3.1 Keto/Enol mechanism: Photonic perspective

Consider $CH_2=CHOH \Leftrightarrow CH_3\text{-}CHO$ to recapitulate some ideas. 1) Both differ in the O-H and C-H entanglements. 2) As noted above Sect.2.2, one can assign local nodal planes with the respective normal vectors making a π/2 angle. There are no symmetry operations allowing geometry mapping between both. In chemistry this fact is used to introduce the concept of reaction coordinate. 3) The keto form does not present a C-C "double" bond instead it locates the feature at the carbonyl "double" bond C-O. Apparently, the number of electrons and nuclei is the same meaning that the *same* elementary materials sustain the quantum states of keto and enol. Both frameworks lie on orthogonal planes, say (x,y)- or (y,z)-. Thus, because keto and enol are different families of electronic states they cannot be transformed via symmetry operations; simple geometric changes are not sufficient, they are not even necessary. To change amplitudes from one set to the other obtains via *reaction coordinates* and physical photonic effects.

How does one breaks entanglement and construct new ones? The question is somewhat incorrectly formulated. One confronts to quantum states not to entangled objects; obviously, entanglement is not an object, it relates to quantum states of particular types. The point is to discover how *amplitudes* at entangled sector can be *shifted* to a non-entangled one, etc. Then looking at the relevant graphs one may indulge the linguistic expression.

Consider a generic 1-partite and a particular related bi-partite system. The second shows up at least one nodal plane compared to the first one; the tree structure differs in one missing link. If both base states share I-frame, superpositions of 1- and bi-partite base states may lead to coherent states. To get free bi-partite elements (fragments) requires a *decoherence transition*.

Start up from two separate fragments; *re-coherence* would open a channel to "move" amplitudes from the bi-partite to the 1-partite state amplitudes. This process would have some likenesses with "bond-forming" or knitting and this would be so if topologic trees enter the discussion. A localized entanglement expresses itself, in chemical language, as a "bond"-like feature.

It remains describing explicit quantum mechanism responsible for either de-coherence or re-coherence to advance the discussion. Consider the generic q-state:

$$(\ldots |j=1m(1)\rangle \otimes |1_\omega\rangle \; |j=1m(1);1_\omega\rangle \ldots |2m(2)\rangle \ldots |3m(3);0_\omega\rangle \ldots |10\,m(10)\rangle \ldots) \bullet$$
$$(\ldots C_{1m(1)\otimes 1_\omega} \; C_{1m(1);1_\omega} \ldots C_{2m(2)} \ldots C_{3m(3)\otimes 0_\omega} \ldots C_{10m(10)} \ldots)^T \rightarrow |q\text{-state}\rangle \qquad (3.1.1)$$

An initial un-entangled state looks like:
$$(\exp(-i\mathbf{k}\cdot\mathbf{x})1_{1m(1)\otimes 1_\omega} \; 0_{1m(1);1_\omega} \ldots 0_{2m(2)} \ldots 0_{3m(3)\otimes 0_\omega} \ldots 0_{10m(10)} \ldots)^T \qquad (3.1.2)$$

Photon-matter entangled state follows after interaction showing a possible structure:
$$(\ldots C_{1m(1)\otimes 1_\omega} \; C_{1m(1);1_\omega} \ldots 0_{2m(2)} \ldots 0_{3m(3)\otimes 0_\omega} \ldots 0_{10m(10)} \ldots)^T \qquad (3.1.3)$$

Propagation from this coherent state to yield (3.1.4) below is to be driven by an external effector accomplishes local entanglement reshuffling:

$$(\ldots C_{1m(1)\otimes 1\omega}\ C_{1m(1);1\omega}\ \ldots 0_{2m(2)}\ldots C_{3m(3)\otimes 0\omega}\ \ldots 0_{10m(10)}\ldots)^T \quad (3.1.4)$$

Note that the free-photon state is now "hooked" as it were to the matter-sustained q-state. This means unavailability to perform semi-classic work; in this context external interactions are implied.

These states are basically energy degenerate with relatively high density-of-states. A q-state propagation, namely relative change of amplitudes patterns, would require low-frequency radiation: e.g. infrared, terahertz, microwaves, radio frequency, etc. Those supplemental photon field energies are "external" to the EN spectra and can be seen either in semi-classic guise or fully quantum physical change. See ref.[2] for an example of the former mechanism.

Spin-orbit coupling (SOC) would allow extension of (3.1.4) to attain a new q-state set; once it is activated one gets for instance spin singlet-triplet mixing:

$$(\ C_{1m(1)\otimes 1\omega}\ C_{1m(1);1\omega}\ \ldots C_{2m(2)}\ldots C_{3m(3)\otimes 0\omega}\ \ldots 0_{10m(10)}\ldots)^T \quad (3.1.5)$$

Analyses of Scheme I show that within enol subspace there is no base state able to opening a pass to keto domain. Yet, state |4,m(4)> displays a spin triplet concerning an O-H element; from a quantum chemical perspective, anti-bonding nature of the base state signals possibility for large O-H fluctuations.

Base states |9, m(9)> and |10, m(10)> in Scheme I as already noticed are simple transposition of chemical equations intending carbon-hydrogen and oxygen-hydrogen bond breaks. They would rather correspond to symmetry broken states. This can be underlined by using (z,x)-plane as reflection element and the base functions would take symmetry adapted ones: take states |9'> and |10'> as a convenient entangled set:

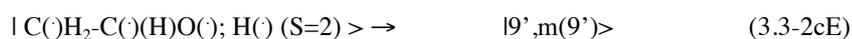

| $C(\cdot)H_2\text{-}C(\cdot)(H)O(\cdot); H(\cdot)\ (S=2) > \rightarrow$ | $|9',m(9')>$ | (3.3-2cE) |

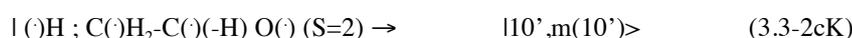

| $(\cdot)H\ ;\ C(\cdot)H_2\text{-}C(\cdot)(\text{-}H)\ O(\cdot)\ (S=2) \rightarrow$ | $|10',m(10')>$ | (3.3-2cK) |

Yet, these bases still target chemically biased base (local) states. A further transformation yields what may be physically sounder:

$$|11, m(11)> = 1/\sqrt{2}\ (|3.3\text{-}2cE > + |3.3\text{-}2cK >) \quad (3.1.6)$$

$$|12, m(12)> = 1/\sqrt{2}\ (|3.3\text{-}2cE > - |3.3\text{-}2cK >) \quad (3.1.7)$$

The H-atom state entangles to the same radical-states though with intended interactions indicated by the semicolon position above, Cf. (3.2-cE) and (3.3-22K). But for states |11, m(11)> and |12, m(12)> the labels K and E are superfluous in the rotated base; they are no longer good quantum numbers. If one wants to describe semi-classically these base states state then (3.1.6) corresponds to delocalized hydrogen atom-states with equal weight for the

keto- and enol-like components: a sort of H-atom "cloud" that signals base states, i.e. *possibility clouds*; this implies an infinity of collision situations in agreement with the new perception of quantum states. Base state (3.1.7) is obviously orthogonal to (3.1.6) and will mediate amplitude changes. In other words, these base states would quantum mechanically sustain a Keto/Enol conversion.

A transition producing the linear superposition between states $|11, m(11)\rangle$ and $|12, m(12)\rangle$ yields non-zero amplitudes for instance only at $|3.2\text{-}2cE\rangle$ which is enol entangled. Semi-classically it suffices to break this bond interaction to put amplitudes at $|9, m(9)\rangle$ which is a fragment state in this case a free hydrogen state. Put simply, entanglement is reverted. But this is not what we are looking after. The amount of energy required is much too large. If one wants to get keto from enol response or vice versa it is not necessary to climb that much in the energy ladder. There is no population implied, only coherent states.

These new base states, namely $|11, m(11)\rangle$ and $|12, m(12)\rangle$, correspond to a 1-partite case, and the elements appearing belong to a coherent q-state; therefore only one I-frame.

Consider q-state (3.1.1) standing as a coherent superposition that includes spin singlet for the vertical excited state, it looks like:

$$(\ldots 0_{1m(1)\otimes 1^\omega} \quad C_{1m(1);1^\omega} \ldots 0_{2m(2)} \quad C_{3m(3)} \ldots 0_{9'm(j')} \ldots 0_{10'm(10')} \ldots)^T \qquad (3.1.2)$$

From Scheme I eq. (3.3-1bE) the spin function is symmetric to label permutations so that it is the space function that must be anti-symmetric; namely an "anti-bonding" situation. But amplitude different from zero at $C_{2m(2)}$ opens opportunities to activate excitations using probes targeting this one as root state, for example $|7\ m(7)\rangle$. In this case Keto subspace becomes accessible. By putting amplitude different from zero at the carbonyl function the process would take the desired direction, yet, there is need to engage the methyl functionality as well, i.e. –$CH_3$. Here, C-center sustains three local entanglements.

This is the role to be played by base state $|9'\ m(9')\rangle$ where quantum numbers for the methyl functionality are engaged. Note that the primed states are coherent superpositions of the unprimed bases found in Scheme I. Only under such conditions can one explore the possibilities for interactions with the remaining base states. This is a quantum mechanical requirement.

So far no other than states (4.1a,b) have pop up as decoherence-window states. Though these ones can be constructed by breaking coherence shown for those states sustained by base states $|9'\ m(9')\rangle$ and $|10'\ m(10')\rangle$. Induce transitions between $|3.3.3\text{-}Tcloud+\rangle$ and $|3.3.3\text{-}Tcloud\text{-}\rangle$ to produce a first result where either amplitudes at $|3.3.2\text{-}cE\rangle$ or at $|3.3.2\text{-}cK\rangle$ could be different from zero. These symmetry broken states are those bridging keto-to-enol transport in Hilbert space

Chemistry with the help of topologic graphs elicits a quantum physics *sustained* by the elementary materials, namely, electrons and nuclei.

### 3.2. Rounding off photonic framework with external sources

So far, the photon field incorporated in the photonic scheme corresponds to entangled states or free fragments in their way to collide and consequently there are no free photon states available. Once preparing an initial q-state, in this scheme, it would stay put. From eq.(2.3.1) it follows a Hamiltonian H given as:

$$H = \sum_{jm(j)} E_{jm(j)} |jm(j)\rangle \langle jm(j)| \qquad (3.2.1)$$

With this operator and using excitation operator [2] that acts on the EN q-state projected over the target state reads:

$$\langle jm(j)| \, I(j,j') \, |q\text{-state}\rangle \rightarrow \langle jm(j)| \, \{ \, H \, |jm(j)\rangle \langle j'm(j')| - |jm(j)\rangle \langle j'm(j')| \, H \, \} \, |q\text{-state}\rangle \rightarrow$$
$$(E_{jm(j)} - E_{j'm(j)}) \, C_{j'm(j')} \qquad (3.2.2)$$

The response is modulated by the value of the amplitude affecting the root state irrespective of the amplitude at the target state. Bohr's mapping $(E_{jm(j)=0} - E_{j'm(j)=0}) = (h/2\pi)\omega$ develops a link between laboratory and abstract worlds; the symbol $\omega$ stands for an appropriate frequency of an EM radiation.

Entangled with specially tailored quantized photon fields, the composite base set permits descriptions of time evolution if *external* time-dependent probes (sources) operates at the laboratory level on the photon-matter entangled quantum states. The activated initial state in a one-photon process would read: $(\ldots 0_{jm(j)\otimes 1\omega} \; 1_{jm(j);1\omega} \ldots \; 0_{j'm(j');0\omega} \ldots 0_{j'm(j')\otimes 0\omega} \ldots)^T$. Internal coupling operators $\hat{V}_{int}$ may mix singlet-to-triplet spin spaces; this operators ought to incorporate external ones if necessary.

Let $\hat{V}(Y_{lab-probe}, X_{q-syst}; t)$ be the operator coupling a q-state to a spatially located device in lab-space. The partite base states are identified as $\langle \mathbf{y}_1, \ldots \mathbf{y}_\nu \, |\mu_{k1\ldots}\mu_{k\nu}\rangle \otimes \langle \mathbf{x}_1, \ldots, \mathbf{x}_m \, |\phi_{\mathbf{k}'1\ldots k'm}\rangle$ with varied partitioning in both sectors; each sector display specific I-frames amounting to lab-space localization. From the set of all possibilities one selects those relevant to the chosen experimental device(s). Consider the case where configuration spaces are not altered by the measurement opening a window to analyze simple setups.

At the point in laboratory space where an interaction takes place the resultant q-state projected in **x**-space with external labels identifying the place where information was reshuffled cannot but be given by a new set of amplitudes reflecting the measuring interaction.[35] The novelty lies in the inclusion of I-frames locations associated to the positions of the measuring devices. This is possible to the extent amplitudes over partite base states change I-frame reference during the process. In this sense the approach would be semi-classic, as it should.

Such entangled q-states result from preparation protocols. If by bringing photon states that would entangle definite electronic states (the way we put is in eq. (2.4.1)) *no time-evolution* would be effective: the photon states are just information. Viewed from a semi-classic perspective the "photon" lost its time-dependent field by being "soaked" as it were with the matter-sustained q-states. This is a particular feature of the present scheme.

Once the initial q-state is prepared, the ingoing photon amplitudes are assigned zero-value; the base set gathers all information of spectra, Cf. eq.(2.4.1) and the quantum state sustained by the elementary materiality is given in eq.(2.4.2). Probing a q-state amounts to identify all non-zero amplitudes. If $C_{j'm(j')}$ equals to zero there cannot be a response even if the radiation frequency is appropriate. Energy levels featuring non-zero amplitudes can act as root state of new spectral series; and these latter can be used for identification purposes. The source of external EM radiation should contain the characteristic frequencies.

The procedure so far permits reckoning (all) possibilities a system may show up. Physical interactions are required to link abstract to laboratory spaces; they convey amplitudes reshuffling. Possibilities must be introduced that include quantum states to be measured and those corresponding to probing quantum states. Quantum states for quantum measurements [5] are called into the picture rounding off the photonic framework. Events signaling energy and/or angular momenta exchanges are those leaving behind a physical imprint in laboratory space. [35] The imprints (or clicks) are countable events; they are the support for probability calculations.

Thus, physical interaction seen from the q-state system produces quantum state changes that otherwise would not be detected. This would be a typified catalytic effect. Cavities found in *ionic liquids* are less known cases deserving further studies that would set up interactions capable of acting as catalytic effectors.

Keto ground EN state locates below enol ground EN level; there exist a set of keto EN levels that have no enol energy levels under resonance conditions. This is a gap contributing to activation barrier. Coming to a region where enol EN levels can be mixed with excited EN keto levels, possibilities for resonance emerge. And resonance conditions might be fulfilled. The point is that low frequency modes would mediate time evolution via EN energy levels, jm(j). [2] And time dependent processes might be induced with low frequency radiation, namely infrared (IR), Tera-Hertz radiation, microwave (MW), etc. [36]

## 4. Closing remarks

The present quantum methodology views chemical processes as akin to electronic "transitions"; [6] *though* now they involve coherent quantum states (Cf.eq.(2.4.3)) and changes thereof modulated by applied fields, in particular electromagnetic ones. Reaction channels open under external steering, shown by developing finite amplitude at a particular electronic base state designated as "product." The present approach seeks to be consistent with modern quantum technologies to control EN systems with all their promise to manipulate structure and reaction mechanisms via spatial confinement and electromagnetic fields.

The emerging physical picture discussed in this paper reminds on the one hand, the role of Feshbach quantum states for material systems entangled with the radiation field. These resonance states act as gates to change amplitudes at entangled bound base states, as well as exit towards entangled asymptotically separated molecular states, thereby emphasizing the quantum physical nature of chemical bond reshuffling processes. The semi-classic (diabatic) model reinforces the qualitative views. Feshbach quantum state might appear to be a quantum path to describe bond reshuffling in general bond alteration processes. On the other hand, as there is no mechanical tweezers, EM radiations can replace such enforcer; cavities that act as tweezers can be found at active sites from enzymes; and most likely in ionic liquids cavities.

The procedure so far permits reckoning all possibilities a system may show up; a pattern of entanglement not mentioned so far is the transport of O-H entanglement to a –C-H one at the *same* C-center and the concurrent entanglement of oxygen with both C-centers corresponding to a oxy-ethane graph. This type of entanglement reshuffling plays a key role in the electronuclear mechanism in Rubisco.[23]

Moreover, the model attains a quantum mechanical structure that makes it adaptable to the study of quantum impurity cases. Thus, we might be able to wed information coming from advanced quantum chemical sources to, for example, continuous-time Monte Carlo methods to study quantum impurity models.

Note that among the steps required to re-construct a quantum picture starting from the semi-classic model the quantization of translational energy plays a key role.

Summing up, theoretical developments and non-routine computations results can hence be directly related to experiment, which gives an encouraging perspective. However, the quantum nature of matter response defies the classical intuition; and one of the reasons for this state of affair originates by interactions with external probing expresses via matter-sustained quantum states and not as a material object (classical physics). Quantum physics by addressing, in principle, all possible quantum states resulting from interactions brings novelties that are difficult to digest and assimilate within current forms of interpreting quantum mechanics. And, mathematical chemistry becomes an essential framework to mediate different levels of presence for material systems.

**Acknowledgments**

The authors are much indebted to Prof. G.A. Arteca for discussions and contributions to early stages of this work. O.T acknowledges discussions with Prof. Erkki Brändas; and he is most grateful to Prof. Eduardo Ludeña for invaluable conversations on Quantum Physics.